\documentclass[12pt,letterpaper]{JHEP3}
\pdfoutput=1
\usepackage{cite}
\usepackage{epsfig}
\usepackage{subfigure}

\graphicspath{{Figures/}}

\title{Strongly Warped BPS Domain Walls}

\author{Ali Masoumi and I-Sheng Yang\\
Institute of Strings, Cosmology and Astroparticle Physics \\
Department of Physics \\
Columbia University, New York, NY 10027, USA
}

\abstract{We present analytical solutions of BPS domain walls in the Einstein-Maxwell flux landscape.  We also remove the smeared-branes approximation and write down solutions with localized branes.  In these solutions the domain walls induce strong (if not infinite) warping.}

\begin{document}

\section{Introduction}

The multiverse is the natural combination of the string theory landscape\cite{BP} and eternal inflation\cite{GutWei83,Linde,GonLin87}.  In the multiverse picture, our universe is not unique, but just one of the $10^{O(100)}$ vacua mutually connected by domain walls (mostly formed by quantum tunnelings).  Therefore these domain walls play important roles in the multiverse theory.  In particular, since our own universe is connected to the multiverse by a domain wall (or several domain walls), its property might be related to some cosmological observables.

\begin{figure}
\begin{center}
\includegraphics[width=14cm]{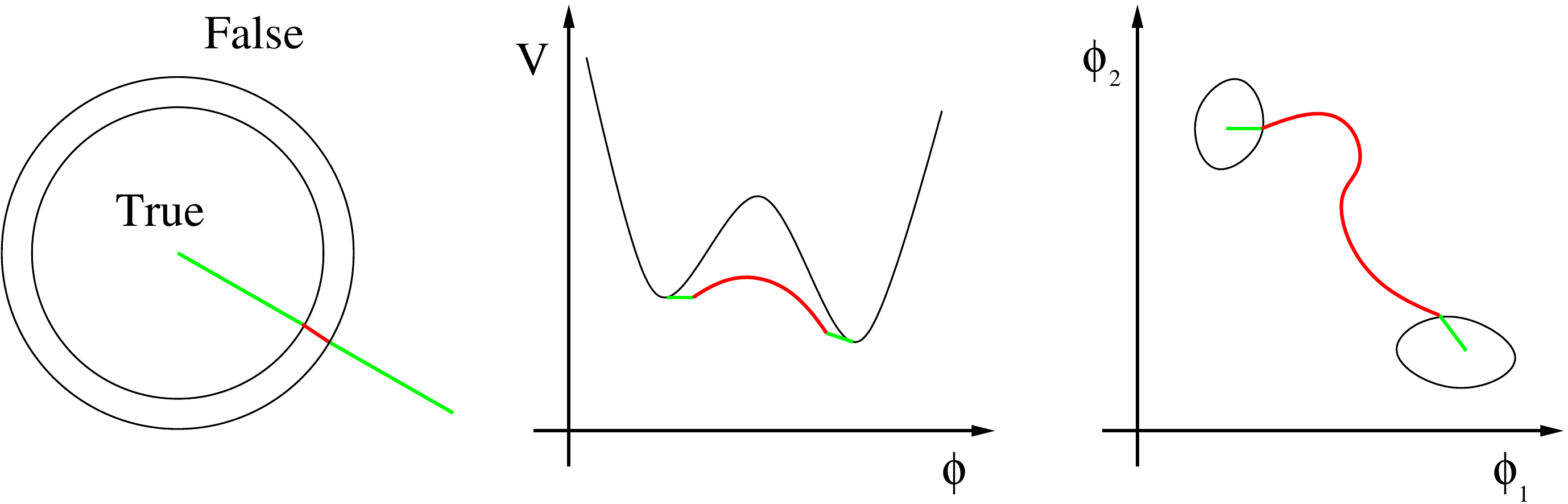}
\caption{The leftmost figure is (part of) a CDL instanton solution.  It has a small bubble of true vacuum embedded in the false vacuum.  Within the thin boundary between two regions, the field has to interpolate from one vacuum to the other (the red segment).  This interpolation is easily derived from the potential barrier if there is only one field.  With multiple fields, it becomes technically nontrivial to find the path.
\label{fig-CDL}}
\end{center}
\end{figure}

Initially the domain walls were not among the major excitements of the string landscape, mostly because it is a much older topic.  In a Coleman-deLuccia instanton\cite{CDL} solution, the domain wall is the interpolating region between two vacua.  In the classic example with only one scalar field, it is just how the field go through the potential barrier, as depicted in Fig.\ref{fig-CDL}.  There does not seem to be much more to say about it.

Cvetic and Soleng\cite{CveSol94} were the first to worry about the fact that the string landscape has multiple fields instead of one, so maybe the story is not as simple as we believed.  Recently it has become clear that a domain wall can exhibit a much richer structure in theories with multiple fields\cite{Yan09,AguJoh09a,AhlGre10,BroDah11}.  Basically, the single field example is the only special case where there is no need to find a ``path'', as shown in Fig.\ref{fig-CDL}.  With multiple fields, it is highly nontrivial to find the right path, which is the source of many interesting physics. 

More importantly, multiple fields are not mere technical complications.  Between two vacua, the need to find a path indeed makes the problem harder.  However in the landscape there are $C^{10^{O(100)}}_{2}$ possible transitions between pairs of vacua, we should not look for explicit paths anyway.  Instead we should look for general rules.  Interestingly, there are increasing evidences that tunneling paths in a multifield phase space do follow some general rules.

The tunneling path tends to go near special points/directions.  The first example is near decompactification.  If one of the fields represents the overall volume of the compact extra dimensions, for example the universal Kahler moduli in type IIB string theory, then the tunneling path tends to make a detour toward where the extra dimensions get large \cite{AguJoh09a}.  The second example is recently shown in \cite{AhlGre10} that the tunneling path goes through a strongly warped region near the conifold point of the field space.

Although the numerical recipe that found the multifield tunneling paths is quite solid, it only works in the thin wall approximation without gravity.  Furthermore, as a general disadvantage of numerical methods, the physical meaning of the results is not always clear.  

The framework of \cite{Yan09} was trying to circumvent these problems, at least in the Einstein-Maxwell model\cite{FreRub80,RanSal82} which recieved a lot of recent attention as a toy model of the string landscape.  Using the duality between a CDL instanton and a Swinger pair production\cite{Sch51}, the ``main'' field of the tunneling is replaced by charged branes with zero thickness.  The other fields manifest as the geometry of compact extra dimensions.\footnote{Note that this is a thin-wall approximation in the higher dimensional theory, which is different from the traditional thin wall.  Since the reaction of the extra dimensional geometry to the thin brane is also part of the domain wall in the effective lower dimensional theory, there is a finite thickness given by the exact solution.}  The entire problem of finding the correct instanton solution becomes a GR problem of finding the correct geometry with given charged branes.

From this point of view, thick wall and gravitational effects are nothing special and can be exactly included.  Sometimes they are not just corrections, but dramatically alter the physical conclusion.  As seen in the case of ``giant leaps'' in the Einstein-Maxwell model\cite{BroDah10}, an effect quite convincing in the thin wall approximation disappeared under a more exact treatment\cite{BroDah10a} in this framework.

Of course there is always a catch.  Generally, finding solutions in GR strongly relies on symmetry anzarts.  For the Einstein-Maxwell model it means we have to maintain the spherical symmetry of the compact dimensions.  Although the charged branes should be points on the compact sphere, they are treated as uniformly everywhere.  This is usually called the smeared branes approximation.

It is important to go beyond smeared branes for at least two reasons.  First as argued in \cite{BlaSch09}, localized branes in a fixed geometry approximation implies classical transitions\cite{EasGib09,GibLam10,JohYan10} when domain walls collide.  Also as argued in \cite{AhlGre10}, localized branes can induce strong warping.

In this paper we will go further under the framework of \cite{Yan09}.  In Sec.\ref{sec-geometry} we demonstrate a family of exact analytical solutions coming from the generalized Reissner-Nordstr\"om metric.  In Sec.\ref{sec-bps} we show that these solutions are BPS domain walls.  Here BPS just means the general property that the action can be broken into complete square terms.

Finally, in Sec.\ref{sec-warp} we show that this family of solutions is simple enough to survive less symmetry.  We remove the smeared branes approximation and show that the extra dimension becomes extremely warped on the domain wall.  This finding echoes the result of \cite{AhlGre10} and provides a way to include geometric backreaction to the scenerio in \cite{BlaSch09}.

It should be noted that all domain walls in this paper are BPS ones.  Strictly speaking, they do not correspond to any tunneling between vacua becasue you cannot build finite action instantons for them.   We cannot provide any proof at this stage, but we believe that the special features, like strong warping, are also presence in a more general family of domain walls, which can lead instanton solutions.

\section{Shell of Branes in the Extremal Geometry}
\label{sec-geometry}

It is well known that Reissner-Nordstr\"om metric can be generalized to any $D$ dimensions with $q$-form fluxes.  In the extremal limit, the metric is 
\begin{eqnarray}
ds^2 &=& f(r)^{\frac{2}{p+1}}(-dt^2+dy_i^2)+f(r)^{-2}dr^2+r^2d\Omega_q^2~,
\label{eq-BH} \\
f(r) &=& 1-\left(\frac{r_*}{r}\right)^{q-1}~.
\end{eqnarray}
Here $D=p+q+2$, $i$ runs from $1$ to $p$, and $r_*$ is the extremal horizon of a black $p$-brane. 

It is also well known that the near horizon limit of the above metric is an $AdS_{p+2}\times S_q$ compactification.  More specifically, it is the flat slicing of such $AdS$ space.
\begin{equation}
ds^2 = e^{2\rho/R_{\rm AdS}}(-dt^2+dy_i^2)+d\rho^2 + r_*^2d\Omega_q^2~.
\label{eq-AdS}
\end{equation}
Note that the $t$ and $y_i$ here are trivially rescaled from those in Eq.~(\ref{eq-BH}), and the $AdS$ radius $R_{\rm AdS}$ is related to the radius of compactified $S_q$ by 
\begin{equation}
R_{\rm AdS}=\frac{p+1}{q-1}r_*~.
\end{equation}

The term ``near horizon limit'' sometimes misled people to think that it is some sort of an approximation.  That is untrue as emphasized in \cite{CarJoh09a}.  Both Eq.~(\ref{eq-BH}) and Eq.~(\ref{eq-AdS}) are exact solutions to Einstein and Maxwell equations of the same total flux.

Such property turns out to be fruitful for physical intuitions.  Practically anything you can do to one metric, you can do it to the other.  One mathematical construction then provides two different physical pictures.  We will exploit this property and demonstrate how an ``extremal shell star'' in the blackbrane geometries corresponds to ``vacuum interpolation'' geometries.

 \subsection{An Extremal Shell}

The asymptotic form of the Reissner-Nordstr\"om metric, Eq.~(\ref{eq-BH}), does not necessarily imply a black $p$-brane in the center.  It can be any extremally charged $p+1$ dimensional object.  The simplest example is a uniformly charged shell.   The interior of such shell will be charge free, therefore a piece of the $D$ dimensional Minkowski space,
\begin{equation}
ds^2 = -dt^2 + dy_i^2 + dr^2 + r^2d\Omega_q^2~,
\label{eq-mink}
\end{equation} 
while the exterior is described by Eq.~(\ref{eq-BH}).

Parameters of the shell can be determined by the Isreal junction condition\cite{Isr66}.  First we calculate the change in extrinsic curvature at the matching radius, $r=\bar{r}$.
\begin{eqnarray}
\Delta K_t^t = \Delta K_y^y &=& \frac{f'(\bar{r})}{p+1} 
= \frac{q-1}{p+1}r_*^{q-1}\bar{r}^{-q}~, \nonumber \\
\Delta K_\Omega^\Omega &=& \frac{f(\bar{r})-1}{\bar{r}} 
= -r_*^{q-1}\bar{r}^{-q}~.
\end{eqnarray} 
Next we specify the material of the shell.  By the symmetry of the problem, in the extended $p$ dimensions the pressure must be the opposite of the energy density, which is the total charge(mass) devided by the surface area proportional to $\bar{r}^q$.
\begin{equation}
\sigma_t^t = \sigma_y^y = \frac{C}{\bar{r}^q}~.
\end{equation} 
Without adding extra ingredients to the theory, these charges only interact through their flux lines which are orthogonal to the shell.  Therefore within the $q$ dimensions of the shell, the pressure is zero.
\begin{equation}
\sigma_\Omega^\Omega=0~.
\end{equation}

Finally the junction condition demands
\begin{eqnarray}
\Delta K_t^t-Tr\{\Delta K\} &=& \sigma_t^t~, 
\label{eq-shell1} \\
\Delta K_\Omega^\Omega-Tr\{\Delta K\} &=& \sigma_\Omega^\Omega~.
\label{eq-shell2}
\end{eqnarray}

Note that we did not start from the most general form of the junction condition. We demanded from the beginning that the matching is static, the boundary is always at $\bar{r}$.  This usually gives us discrete solution(s) as we solve the Eq.~(\ref{eq-shell1}) and (\ref{eq-shell2}) for $\sigma_t^t$ and $\bar{r}$, which indicates the special place(s) where the forces balance out.

Here, Eq.~(\ref{eq-shell2}) is always true and Eq.~(\ref{eq-shell1}) says
\begin{equation}
\frac{p+q}{p+1}r_*^{q-1}=C~.
\label{eq-charge}
\end{equation}
As long as the total charge(mass) agrees with the asymptotic metric, this shell can be anywhere outside the horizon, $\bar{r}>r_*$.  This comes from the fact that for extremally charged objects, gravitational attraction and charge repulsion exactly cancel each other.  There is no net force between any pair of branes on the shell, so they are happy to be anywhere.

 \subsection{A Wall of Nothing}
 \label{sec-won}
Since the charged shell can be anywhere, we can make it approach the horizon, $\bar{r}\rightarrow r_*$.  That actually means matching to the $AdS_{p+2}\times S_q$ geometry.  Namely, the interior is still the Minkowski space, Eq.~(\ref{eq-mink}), but the exterior is a piece of Eq.~(\ref{eq-AdS}).  The matching radius $\bar{r}=r_*$ is required since in Eq.~(\ref{eq-AdS}) the radius of $S_q$ can only take this value.

The extrinsic curvature in this case is slightly different.
\begin{eqnarray}
\Delta K_t^t = \Delta K_y^y &=& R_{\rm AdS}^{-1} 
= \frac{q-1}{p+1}r_*^{-1}~, \nonumber \\
\Delta K_\Omega^\Omega &=& -r_*^{-1}~.
\end{eqnarray}
Nevertheless, the junction condition reaches an identical conclusion that the domain wall charge(mass) is given by Eq.~(\ref{eq-charge}), and the matching can be in an arbitrary position $\bar{\rho}$.  The resulting spacetime is
\begin{eqnarray}
ds^2 &=& e^{2\rho/R_{\rm AdS}}(-dt^2+dy_i^2)+d\rho^2 + r_*^2d\Omega_q^2~,
\ \ \ \ {\rm for} \rho>\bar{\rho}~,\nonumber \\
ds^2 &=& -dt^2 + dy_i^2 + dr^2 + r^2d\Omega^2~, \ \ \ \ {\rm for}
\ \ \ r<r_*~.
\label{eq-won}
\end{eqnarray}

\begin{figure}
\begin{center}
\includegraphics[width=12cm]{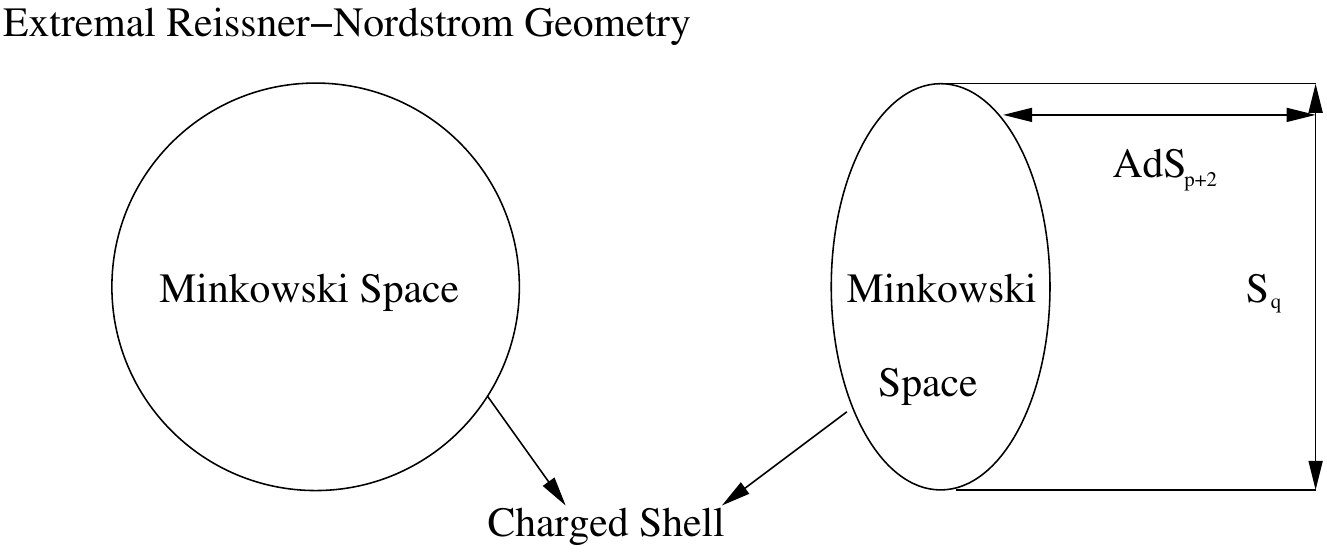}
\caption{Before reaching the horizon of a Reissner-Nordstr\"om geometry, one can replace the center by a piece of Minkowski space and a charged shell.  Similarly, the same Minkowski space and charged shell could be attached to an $AdS_{p+2}\times S_q$ compactification and forms the ``wall of nothing'' geometry.
\label{fig-won}}
\end{center}
\end{figure}

Although mathematically there is nothing new here\footnote{With $p=3$, $q=5$, this is the well known example in string theory that a stack of D3 branes makes an $AdS_5\times S_5$.}, we should take a closer look at the geometry from the lower dimensional $AdS_{p+2}$ point of view.  As depicted in Fig.\ref{fig-won}, this $p+2$ dimensional spacetime stops at $\bar{\rho}$ because the extra dimension $S_q$ shrinks to zero smoothly\footnote{Strickly speaking, the geometry in Eq.~(\ref{eq-won}) is not smooth as it contains a codimension 1 kink at the charge shell.  However it is straight forward to smooth out that shell into a smooth charge distribution.  This will be clear in Sec.\ref{sec-warp}.  One can also look at the smooth solutions in\cite{BlaShl10,BlaRam10}.}in the Minkowski region.  This is the general behavior as the boundary of a bubble of nothing\cite{Wit81}.  In our case it is an infinitely extended flat object, so it is a wall of nothing.  This is also the extremal case of the critical bubble between a bubble of nothing and a bubble from nothing\cite{BlaRam11}.

\subsection{A Domain Wall Between Two Vacua}
\label{sec-wall}

It was suggested in\cite{Yan09} and later shown explicitly in \cite{BlaShl10,BlaRam10,BroDah10b} that from the point of view of multple vacua in the flux-compactified Einstein Maxwell theory, ``nothing'' is like the lowest vacuum with zero flux.  A bubble of nothing is just a special case of decaying from one vacuum to another.

Following the same idea, we can generalized the wall of nothing found in Sec.\ref{sec-won}, to a domain wall interpolating between two $AdS_{p+2}\times S_q$ vacua.  First consider the matching of two extremal Reissner-Nordstr\"om solution with different horizon radius.
\begin{eqnarray}
ds^2 &=& f(r)^{\frac{2}{p+1}}(-dt^2+dy_i^2)+f(r)^{-2}dr^2+r^2d\Omega_q^2~,
\nonumber \\
f(r) &=& 1-\left(\frac{r_1}{r}\right)^{q-1}~, \ \ \ \ 
{\rm for}\ \ \  r>\bar{r}~, \\
&=& 1-\left(\frac{r_2}{r}\right)^{q-1}~, \ \ \ \ 
{\rm for}\ \ \ r<\bar{r}~. \nonumber
\end{eqnarray}
We demand that $r_1>r_2$ so the bigger blackbrane is outside, and $\bar{r}>r_1$ so the matching happens outside its horizon.  It is straight forward to apply the junction condition to see that the charge(mass) of the shell is just the difference of the blackbrane charges.
\begin{equation}
C = \frac{p+q}{p+1} (r_1^{q-1}-r_2^{q-1})~.
\end{equation}
And again this shell can be at any $\bar{r}\geq r_1$.

Pushing the shell to $\bar{r}=r_1$ is the same as replacing the outside of this matching by the $AdS_{p+2}\times S_q$ metric.

\begin{eqnarray}
ds^2 &=& e^{2\rho/R_1}(-dt^2+dy_i^2)+d\rho^2 + r_1^2d\Omega_q^2~,
\ \ \ \ {\rm for} \ \ \  \rho>\bar{\rho}~, 
\label{eq-wall} \\
ds^2 &=& \left[1-\left(\frac{r_2}{r}\right)^{q-1}\right]^{\frac{2}{p+1}}
(-dt^2+dy_i^2)+\left[1-\left(\frac{r_2}{r}\right)^{q-1}\right]^{-2}dr^2
+r^2d\Omega_q^2~,\ \ {\rm for} \ \ r<r_1~. \nonumber
\end{eqnarray}

\begin{figure}
\begin{center}
\includegraphics[width=12cm]{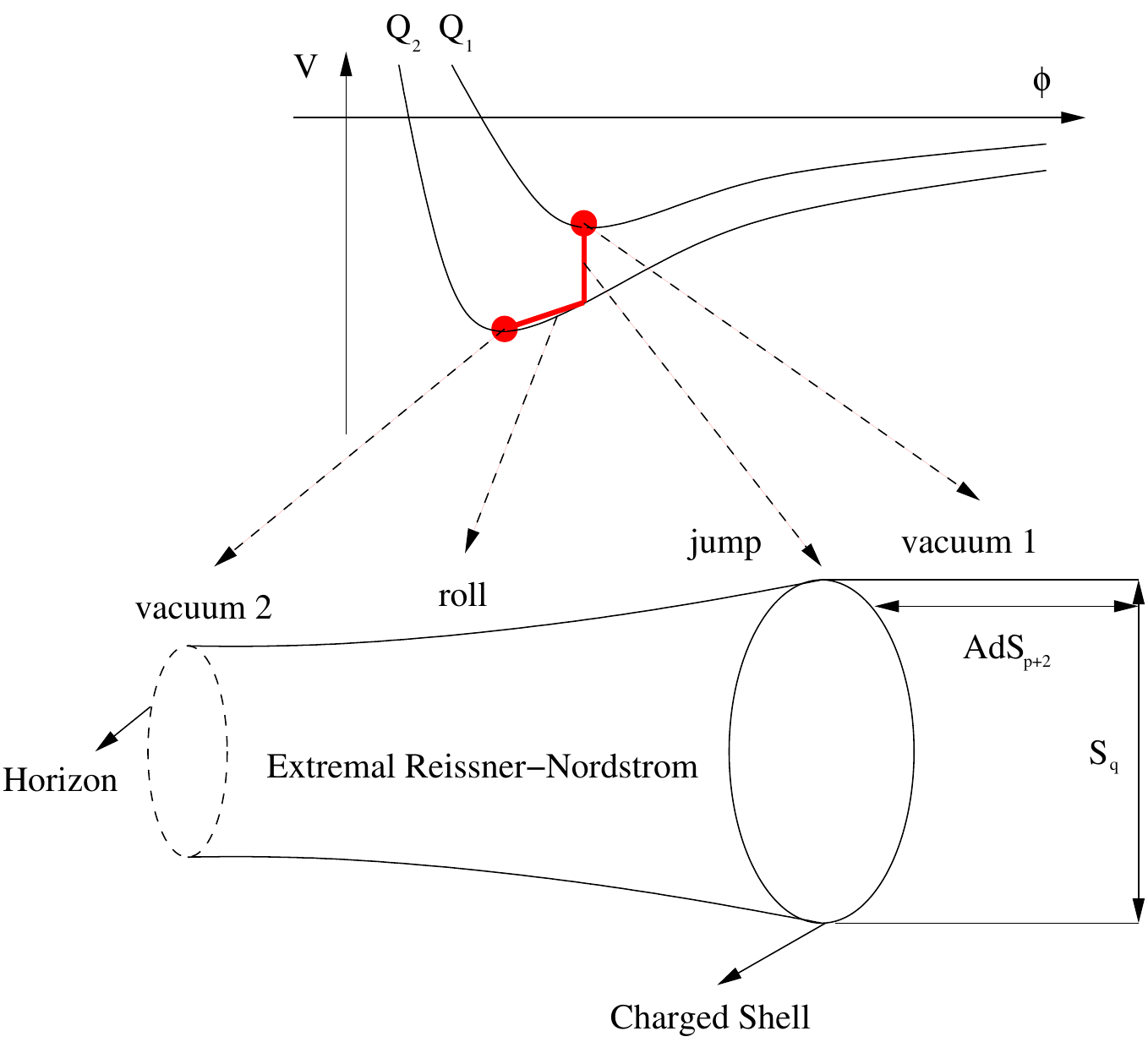}
\caption{The lower half shows a shell of charges that connects a piece of Ressner-Nordst\"orm geometry to $AdS_{p+2}\times S_q$.  The other end of the Ressner-Nordst\"orm geometry approaches its horizon, which is another $AdS_{p+2}\times S_q$.  Therefore this is an interpolation between two vacua.  The upper half shows the corresponding $(p+2)$ dimensional description, where the radion field jumps out of vacuum 1 to the potential of vacuum 2, and then rolls down to vacuum 2. \label{fig-wall}}
\end{center}
\end{figure}

This is a piece of $AdS_{p+2}\times S_q$ in the flat slicing down to an arbitrary $\bar{\rho}$.  The $AdS$ radius is related to the radius of $S_q$ by $R_1 = \frac{p+1}{q-1}r_1$.  Beyond $\bar{\rho}$ the size of $S_q$ monotonically drops from $r_1$ to $r_2$ as we move closer to the horizon of the Reissner-Nordstr\"om  metric, and approaches the ``near horizon'' $AdS_{p+2}\times S_q$ with $R_2=\frac{p+1}{q-1}r_2$.  See the illustration in Fig.\ref{fig-wall}.

\section{A BPS Domain Wall}
\label{sec-bps}

We can also visualize this interpolation between two vacua from a different prospective.  By dimensional reduction, the $(p+q+2)$ dimensional theory with $q$ form flux becomes a $(p+2)$ dimensional theory with a scalar field.  Following the convention in\cite{CarJoh09}, we parametrize the radius of $S_q$ by the radion field $\phi$,
\begin{equation}
r = M_D^{-1}\exp\left[\sqrt{\frac{p}{q(p+q)}}\frac{\phi}{M_{p+2}}\right]~.
\end{equation}  
Here $M_D$ is the Planck mass in $D=p+q+2$ dimensions, and $M_{p+2}$ is the Planck mass in $(p+2)$ dimensions.  They are related by the area of unit $S_q$.
\begin{equation}
M_{p+2}^p = M_D^p {\rm Vol}(S_q)~.
\end{equation}

The $p+2$ dimensional effective action is
\begin{equation}
S = \int d^{p+2}x\sqrt{-g}\left[
\frac{M_{p+2}^p}{2}{\cal R}
-\frac{M_{p+2}^{p-2}}{2}g^{\mu\nu}\partial_\mu\phi\partial_\nu\phi
-V(\phi) \right]~,
\label{eq-S}
\end{equation}
where 
\begin{eqnarray}
V(\phi) &=& \frac{M_{p+2}^p M_D^2}{2}\bigg[
-q(q-1)\exp\left(-2\sqrt{\frac{p+q}{pq}}\frac{\phi}{M_{p+2}}\right)
\nonumber \\
&+& \frac{Q^2}{2}\exp\left(-2(p+1)\sqrt{\frac{q}{p(p+q)}}
\frac{\phi}{M_{p+2}}\right)
\bigg]~.
\label{eq-V}
\end{eqnarray}
The unitless charge $Q$ in their convention is related to the charge $C$ defined in Eq.~(\ref{eq-charge}) by
\begin{equation}
Q = C M_D^{q-1} \sqrt{\frac{2(q-1)(p+1)}{p+q}}~. 
\label{eq-charge1}
\end{equation}

Different $Q$s provide different effective potentials for the radion field $\phi$.  As described in\cite{Yan09}, the $(p+q+2)$ dimensional geometry in Eq.~(\ref{eq-wall}) has a dual $(p+2)$ dimensional description as shown in Fig.\ref{fig-wall}.  The charged shell corresponds to where $\phi$ jumps out of vacuum 1 onto the potential given by $Q_2$, and then follows that potential and rolls down to vacuum 2.

Usually, this rolling process coupled to gravity is too hard to solve analytically.  Here we see that the extremal Reissner-Nordstr\"om geometry provides an exact analytical solution.  This is not an accident.

This analytical form of the metric, Eq.~(\ref{eq-BH}), relies on two important ingredients.  The black brane has a planar symmetry, and the $D$ dimensional cosmological $\Lambda_D=0$.\footnote{If Eq.~(\ref{eq-BH}) can be generalized to include nonplanar symmetry and/or nonzero $\Lambda_D$, then there would be more general solutions.  But we do not know of such generalizations other than in $D=4$.  Including a dynamical dilaton may help to have curved branes\cite{CheBer11}, but as far as we know no analytical solutions are provided there either.}  Consequently in the compactified theory, the interpolation also has planar symmetry, and only between $AdS_{p+2}$ vacua (instead of Minkowski or $dS_{p+2}$).

So what we have is an exact solution coupled to gravity that is a planar interpolation between two $AdS$ vacua.  It sounds like the well-known BPS domain wall, and indeed it is one.  Note that ``BPS'' here is not necessarily related to any supersymmetry, it is just the simple fact that the action, Eq.~(\ref{eq-S}), can be written as the integral of the sum of complete square terms and a boundary term\footnote{This is sometimes known as the fake supergravity\cite{FreNun03}.  Na\"ively one can argue that our model works in any dimensions and supergravity is limited to less than 11 dimensions, so they are definitely not related.  However, the dimension limit for supergravity is in the quantum level.  As a classical theory it is totally fine in any dimension, as long as we introduce higher spin components into the supermultiplet.  It might still be true that our solution is some SUSY preserving configuration, just with a more extended supermultiplet.}.  This means the usually second order equations of motion becomes first order and much easier to solve.  The necessary and sufficient condition for a $(p+2)$ dimensional BPS solution is that the potential $V$ can be written as
\begin{equation}
V(\phi) = \left(M_{p+2}\frac{dW(\phi)}{d\phi}\right)^2-
\left(\frac{p+1}{p}\right)W(\phi)^2~.
\label{eq-BPS}
\end{equation}
From Eq.~(\ref{eq-V}), we get the superpotential $W$ as
\begin{eqnarray}
W(\phi) &=& \frac{M_{p+2}M_D}{\sqrt{2}}\bigg[
q \exp\left(-\sqrt{\frac{p+q}{pq}}\frac{\phi}{M_{p+2}}\right)
\nonumber \\
&+& C M_D^{q-1} \exp\left(-(p+1)\sqrt{\frac{q}{p(p+q)}}
\frac{\phi}{M_{p+2}}\right)
\bigg]~.
\end{eqnarray}
Here we also translated the charge by Eq.~(\ref{eq-charge1}).

\section{A Domain Wall That Warps the Extra Dimensions}
\label{sec-warp}

One can write down even more general solutions of the Einstein-Maxwell model in the following way,
\begin{eqnarray}
ds^2 = U^{\frac{-2}{p+1}}( -dt^2 &+& d\vec{y}^2 ) + U^{\frac{2}{q-1}}d\vec{x}^2~,
\label{eq-multi}
\\
\nabla_x^2 U(x) &=& 0 ~. 
\end{eqnarray}
Here $\vec{y}$ and $\vec{x}$ are vectors in $p$ and $(q+1)$ dimensional flat spaces respectively, and $\nabla_x^2$ is the laplacian operator for $\vec{x}$ only.

For example, the ``point charge potential'',
\begin{equation}
U =\frac{r_*^{q-1}}{|\vec{x}|^{q-1}}~,
\end{equation}
corresponds to a compactified solution $AdS_{p+2}\times S_q$ with $S_q$ radius $r_*$, as in Eq.~(\ref{eq-AdS}), and
\begin{equation}
U = 1+\frac{r_*^{q-1}}{|\vec{x}|^{q-1}} ~ 
\end{equation}
corresponds to an extremal blackbrane solution with horizon radius $r_*$, as in Eq.~(\ref{eq-BH}).

In both cases, $\vec{x}=0$ is not a singularity but an horizon.  The solution is not singular but it is geodesically incomplete.  The other side of this horizon can be the interior of the extremal blackbrane, or the timelike flat slicings of $AdS_{p+2}\times S_q$.

Here we can use superposition to construct more solutions in a straight forward way.  The ``wall of nothing'', Eq.~(\ref{eq-won}), corresponds to a shell of charge.
\begin{eqnarray}
\rho(\vec{x})&=& r_*^{q-1}\delta(r-|\vec{x}|)~, \\
U(\vec{x}) &=& 
\int \frac{\rho(\vec{x}')}{|\vec{x}-\vec{x}'|^{q-1}}d\vec{x}'^{q+1}
= \frac{r_*^{q-1}}{|\vec{x}|^{q-1}} \ \ {\rm or} 
\ \ \frac{r_*^{q-1}}{r^{q-1}}  \nonumber \\
&~& \ \  \ \ \ \ \ \ \ \ \ \ \ \ \ \ \ \ \ \ \ \ \
{\rm for} \ \ |\vec{x}|>r \ \ {\rm or} \ \ |\vec{x}|<r~. \nonumber
\label{eq-won1}
\end{eqnarray}

Similarly, a BPS domain wall between two $AdS_{p+2}\times S_q$, Eq.~(\ref{eq-wall}), corresponds to a point charge with a shell of charges.
\begin{eqnarray}
\rho(\vec{x})&=& (r_1^{q-1}-r_2^{q-1})\delta(r-|\vec{x}|)~,  \\
U(\vec{x}) &=& \frac{r_2^{q-1}}{|\vec{x}|^{q-1}} + \int \frac{\rho(\vec{x}')}{|\vec{x}-\vec{x}'|^{q-1}}d\vec{x}'^{q+1}
= \frac{r_2^{q-1}}{|\vec{x}|^{q-1}} \ \ {\rm or} \ \ 
\frac{(r_1^{q-1}-r_2^{q-1})}{r^{q-1}}+\frac{r_1^{q-1}}{|\vec{x}|^{q-1}} \nonumber \\
&~& \ \  \ \ \ \ \ \ \ \ \ \ \ \ \ \ \ \ \ \ \ \ \ \ \ \ \ \ \ \ \ \ \ \ \ 
{\rm for} \ \ |\vec{x}|>r \ \ {\rm or} \ \ |\vec{x}|<r~. \nonumber
\label{eq-transsmear}
\end{eqnarray}

Note that when we smear point charges into a surface density in these solutions, the geodesic incompleteness is gone.  Their positions no longer correspond to extremal blackbrane horizons.  These surfaces are really physical charges.  It is then straight forward to smear them even more to be volume densities of charge, and the solution is completely smooth.  

This approach has the advantage that the smeared charge distribution can respect the symmetry of $S_q$, therefore all dynamics are in its size and summarized by the bebavior of the radion field.

However, microsopically, the charge has a natural quantization unit and the smallest charge may have a size much smaller than the size of the extradimensions.  From this point of view we cannot smear the charge around the entire $S_q$, and the domain wall may be better discribed bt the two center BH solution\cite{Bri91}.
\begin{equation}
U(\vec{x}) = \frac{r_2^{q-1}}{|\vec{x}|^{q-1}} + 
\frac{(r_1^{q-1}-r_2^{q-1})}{|\vec{x}-\vec{x}_0|^{q-1}}~.
\label{eq-transpoint}
\end{equation}
Eq.~(\ref{eq-transsmear}) and (\ref{eq-transpoint}) have exactly the same asymptotic behavior at $\vec{x}\rightarrow0$ and $\vec{x}\rightarrow\infty$.\footnote{Technically speaking, in Eq.~(\ref{eq-transsmear}) we are already in the later vacuum when $|\vec{x}|>r$, but we have the freedom to take $\vec{x}\rightarrow\infty$ and it does not make a difference.}  Therefore, they connect the same pair of $AdS_{p+2}\times S_q$ vacua.  Furthermore they both have planar symmetry, so they must have the same tension.  From the $(p+2)$ dimensions point of view, they are identical.  The only difference is that the symmetry of $S_q$ is broken by $\vec{x}_0$, and the metric now depends on an internal coordinate of $S_q$, 
\begin{equation}
\theta = \cos^{-1}\frac{\vec{x}\cdot\vec{x}_0}{|\vec{x}||\vec{x}_0|}~,
\end{equation}
which means warping\cite{RS}.  It is somewhat trickier than the usual case as the metric also depends on one of the noncompact $(p+2)$ coordinates.  It is natural to rearrange the metric as
\begin{eqnarray}
ds^2 = A^{\frac{-2}{p+1}}r^{2\frac{q-1}{p+1}}( -dt^2 &+& d\vec{y}^2 ) + A^{\frac{2}{q-1}}\frac{dr^2}{r^2} + A^{\frac{2}{q-1}}d\Omega_q^2~,
\label{eq-metricwarp}
\end{eqnarray}
and recognize 
\begin{equation}
A(r,\theta) \equiv r^{q-1} U = r_2^{q-1} 
+ (r_1^{q-1}-r_2^{q-1})(1+\frac{|\vec{x}_0|^2}{r^2}
-2\frac{|\vec{x}_0|}{r}\cos\theta)^{-\frac{q-1}{2}}
\end{equation} 
as the dynamical warp factor.  In both assymptotics, $r\rightarrow0$ and $r\rightarrow\infty$, $A$ are constants and the first two terms naturally combines to $AdS_{p+2}$.  In between them, the warping depends on $r$, the coordinate orthogonal to the domain wall.  Therefore we say the warping is dynamically induced by the domain wall.  

\begin{figure}
\begin{center}
\includegraphics[width=8cm]{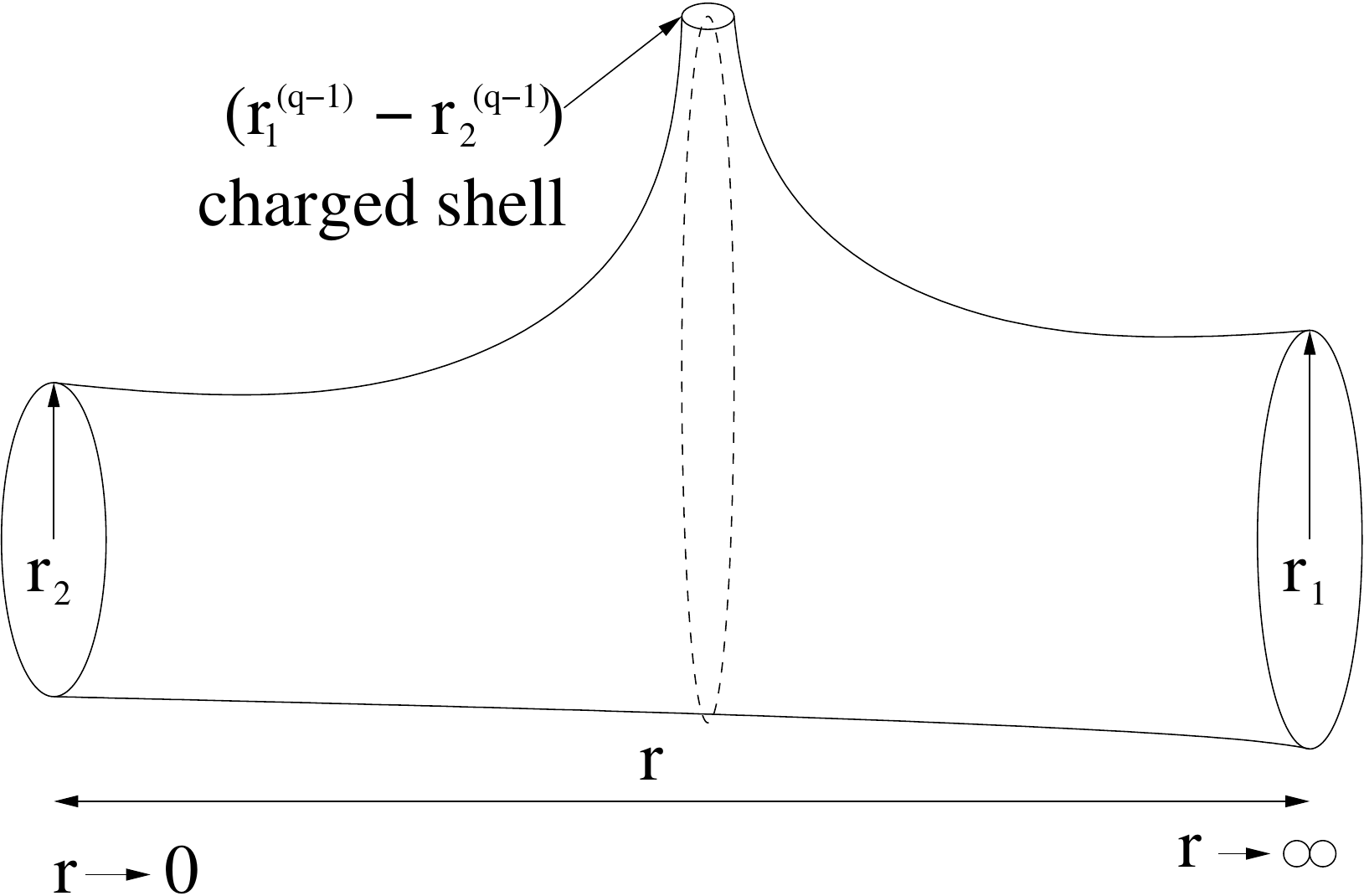}
\caption{The geometry given by Eq.~(\protect\ref{eq-metricwarp}) and (\protect\ref{eq-cutoffwarp}).  To the left as $r\rightarrow0$ it approaches an $AdS_{p+2}\times S_q$ the $S_q$ radius $r_2$.  To the right as $r\rightarrow\infty$ it approaches another compactified vacuum with radius $r_1$.  This transition is mediated by a charge $(r_1^{q-1}-r_2^{q-1})$ which locates at the top tip.  Due to the presence of this charge, the dotted circle is strongly warped though still topologically an $S_q$.
\label{fig-warp}}
\end{center}
\end{figure}

In particular, when $r=|\vec{x}_0|$, we have infinite warping as $\theta\rightarrow0$.  As mentioned earlier, near this point the metric in Eq.~(\ref{eq-metricwarp})is geodesically incomplete.  It is more reasonable to replace the second term of Eq.~(\ref{eq-transpoint}) by some charge distribution instead of a point charge.  We can use a charged shell,
\begin{eqnarray}
\rho(\vec{x})&=& (r_1^{q-1}-r_2^{q-1})
\delta(\varepsilon-|\vec{x}-\vec{x}_0|)~, \nonumber \\
U(\vec{x}) &=& \frac{r_2^{q-1}}{|\vec{x}|^{q-1}} + \int \frac{\rho(\vec{x}')}{|\vec{x}-\vec{x}'|^{q-1}}d\vec{x}'^{q+1}~,
\label{eq-cutoffwarp}
\end{eqnarray}
to make it geodesically complete near $\vec{x}_0$ as a wall of nothing geometry, Eq.~(\ref{eq-won1}).  Far away from $\vec{x}_0$ the small shell is like a point charge, so it is identical to Eq.~(\ref{eq-transpoint}).  We plot the physical picture of this geometry in Fig.\ref{fig-warp} for better understanding.

Note that the domain wall induced strong warping is the same as the result of\cite{AhlGre10}.  It is much more straight forward here to see that such warping is exactly the back reaction of charges on the geometry.  The string theory model in\cite{AhlGre10} started from smearing all branes.  It is quite curious that instead of the smeared brane solutions in Sec.\ref{sec-geometry}, a localized object emerges from the domain wall dynamics.  This may also related to the fundamental difference between smeared and localized sources in string theory\cite{BlaDan10}.

It is also argued that BPS domain walls in \cite{AhlGre10} has to be infinitely warped, similar to our situation before the geodesics incompleteness is cured.  We hope our construction here can help to find an analytical solution of the BPS domain walls in the more complicated model and further understand the warping dynamics.

\acknowledgments 
We thank Brian Greene, Matt Johnson and Erick Weinberg for helpful discussions.  This work is supported in part by the US Department of Energy.

\bibliographystyle{utcaps}
\bibliography{all}
\end{document}